\documentclass[prf]{revtex4-2}

\usepackage{amsmath,amsfonts,amssymb,bm,dsfont,mathtools,xcolor}
\usepackage{graphicx,scrextend}
\usepackage{enumerate}

\DeclarePairedDelimiter{\set}{\{}{\}}

\DeclarePairedDelimiter{\norm}{|\!|}{|\!|}
\DeclareMathOperator{\sign}{sign}

\newcommand{\eps}{\varepsilon}
\newcommand{\grad}{\nabla}

\renewcommand{\d}{\mathbf{d}}

\newcommand{\f}{\mathbf{f}}

\renewcommand{\k}{\mathbf{k}}
\newcommand{\m}{\mathbf{m}}
\newcommand{\n}{\mathbf{n}}
\newcommand{\p}{\mathbf{p}}

\renewcommand{\u}{\mathbf{u}}

\newcommand{\x}{\mathbf{x}}
\newcommand{\z}{\mathbf{z}}

\newcommand{\tu}{\tilde{\mathbf{u}}}

\newcommand{\B}{\mathbf{B}}

\newcommand{\E}{\mathbf{E}}
\newcommand{\F}{\mathbf{F}}

\renewcommand{\H}{\mathbf{H}}

\newcommand{\I}{\mathbf{I}}
\newcommand{\Q}{\mathbf{Q}}

\renewcommand{\S}{\mathbf{S}}
\newcommand{\T}{\mathbf{T}}

\newcommand{\X}{\mathbf{X}}

\newcommand{\tQ}{\tilde{\mathbf{Q}}}

\newcommand{\cB}{\mathcal{B}}

\newcommand{\cE}{\mathcal{E}}

\newcommand{\cI}{\mathcal{I}}

\newcommand{\cK}{\mathcal{K}}
\newcommand{\cL}{\mathcal{L}}

\newcommand{\cS}{\mathcal{S}}

\newcommand{\cU}{\mathcal{U}}

\newcommand{\btau}{\bm\tau}

\begin{document}

\title{Conformations, correlations, and instabilities of a flexible fiber in an active fluid}
\author{Scott Weady$^{1}$}
\author{David B. Stein$^{1}$}
\author{Alexandra Zidovska$^{2}$}
\author{Michael J. Shelley$^{1,3}$}
\date{\today}

\affiliation{$^{1}$Center for Computational Biology, Flatiron Institute, New York NY, 10010, USA\\
$^{2}$Department of Physics, New York University, New York, NY, 10003, USA\\$^{3}$Courant Institute of Mathematical Sciences, New York University, New York, NY, 10012, USA}

\begin{abstract}
Fluid-structure interactions between active and passive components are important for many biological systems to function. A particular example is chromatin in the cell nucleus, where ATP-powered processes drive coherent motions of the chromatin fiber over micron lengths. Motivated by this system, we develop a multiscale model of a long flexible polymer immersed in a suspension of active force dipoles as an analog to a chromatin fiber in an active fluid -- the nucleoplasm. Linear analysis identifies an orientational instability driven by hydrodynamic and alignment interactions between the fiber and the suspension, and numerical simulations show activity can drive coherent motions and structured conformations. These results demonstrate how active and passive components, connected through fluid-structure interactions, can generate coherent structures and self-organize on large scales.
\end{abstract}

\maketitle

\section{Introduction}

Many biological systems involve interactions between active processes and passive, deformable structures. Examples include microorganisms navigating near membranes \cite{Fauci1995,Trouilloud:2008,Dias:2013,Nambiar:2022,Spagnolie2023}, droplets with active interiors \cite{Zwicker:2017,Young2021,Kokot2022}, and mixtures of cytoskeletal filaments and molecular motors \cite{Foster2015,Needleman2017,Berezney2022}. Fluid-structure interactions, mediated by boundary conditions, and steric effects play an important role in many of these systems, giving rise to both local and long-range forces that can drive large-scale organization. Flexible polymers, both active and passive, immersed in liquid crystalline or particle-laden environments provide a rich class of examples. Such systems demonstrate a wide range of phenomenology, including conformational instabilities \cite{Kikuchi2009}, localization and collapse \cite{Harder2014,Mousavi2021}, and modified elasticity \cite{Eisenstecken2017}. 

Chromatin, which is a complex of DNA and histone proteins, is a particularly important and motivating example. During interphase -- the time between two cell divisions -- chromatin is loosely packed and undergoes various processes essential for genome function and maintenance, including transcription, replication, and DNA repair \cite{Alberts2017,Hubner2010,Zidovska2020b}. While chromatin appears disordered during this time, measurements of chromatin displacements in live nuclei show that, on long timescales, chromatin motions are correlated over micron lengths, comparable to the scale of chromosome territories \cite{Zidovska2013}. When ATP is depleted, these long-range correlations do not occur, suggesting activity internal to the nucleus drives its coherent motions \cite{Zidovska2013,Saintillan2018}. 

From a modeling perspective, chromatin is often described as a polymer chain, either active or passive \cite{Ganai2014,Vasquez2016,Hult2017,Liu2018,Shi2018,di2018anomalous,DiStefano2021,Liu2021}. Indeed, simulations of an active polymer immersed in a Newtonian fluid reproduce the activity-driven correlations observed in live nuclei and show that such correlations depend on the nature of microscopic activity \cite{Saintillan2018}. In particular, extensile activity enhances correlated motion while contractile activity merely magnifies random fluctuations. Activity can also lead to coherent configurations of the polymer, including elongation, stiffening, and nematic alignment \cite{Mahajan2022b}. Collections of multiple active polymers demonstrate an even broader range of dynamics such as mixing and segregation, analogous to mixtures of hetero- and euchromatin \cite{Mahajan2022}. 

While activity largely takes place along the chromatin fiber, the surrounding nucleoplasm itself may be active. In particular, the nucleoplasm contains enzymes as well as subnuclear bodies, such as nucleoli, speckles and Cajal bodies, which themselves are sites of active processes. These active particles and processes may contribute to nucleoplasmic flows and chromatin motions in the cell nucleus \cite{Caragine2018,Caragine2019,Zidovska2020b,Zidovska2020a}. In this paper, as an analogy to chromatin in an active nucleoplasm, we develop and analyze a multiscale model of a long flexible fiber immersed in an active environment. Starting from kinetic theory, we describe a continuum model of an active suspension of force dipoles. We then discuss the fiber dynamics and elasticity, including alignment interactions with the underlying medium. In the continuum limit this model admits a steady state, parameterized by the strength of alignment interactions, and linear stability analysis of this state shows the passive fiber can induce an orientational instability in the active fluid which would otherwise not occur. Finally, we perform numerical simulations of the model in spherical confinement, where we find the fiber-active fluid system self-organizes in both its configurations and long-time motion.

\section{Multiscale Model}

\subsection{The active fluid model}

At small length scales where inertia is negligible, the leading order stress contribution of a force-free active particle is a dipole \cite{Batchelor:1970}. Since the  systems we consider are typically on the order of microns, we represent the active environment as a suspension of force dipoles. Here we use kinetic theory, where the suspension is described by a distribution function $\Psi(\x,\p,t)$ that represents the probability of finding a particle at position $\x$ with orientation $\p$ \cite{Doi1986,SS2008}. We consider a suspension of $N$ particles confined in a sphere $\Omega\subset\mathds{R}^3$ of radius $R$ so that $\int_\Omega\int_S \Psi ~ d\p d\x = N$, where $S = \set{\p\in\mathds{R}^3:|\p|=1}$ is the unit sphere of orientations. Assuming the total number of particles is conserved, this distribution function satisfies a Smoluchowski equation,
\begin{equation}
\frac{\partial\Psi}{\partial t} + \grad_x\cdot(\dot\x\Psi) + \grad_p\cdot(\dot\p\Psi) = 0,\label{eq:dPsi/dt}
\end{equation}
where $\grad_x = \partial_{\x}$ is the spatial gradient, $\grad_p = (\I-\p\p)\cdot\partial_{\p}$ is the surface gradient on the unit sphere, and $\dot\x$ and $\dot\p$ are configuration fluxes which describe the microscopic dynamics of a single particle. In the dilute limit, these fluxes are given by
\begin{align}
\dot\x &= \u - D_T\grad_x\log\Psi,\label{eq:xdot}\\
\dot\p &= (\I-\p\p)\cdot\grad\u\cdot\p - D_R\grad_p\log\Psi,\label{eq:pdot1}
\end{align}
where $\u(\x,t)$ is the fluid velocity with the convention $(\grad\u)_{ij} = \partial u_i/\partial x_j$, and $D_T$ and $D_R$ are translational and rotational diffusion coefficients, respectively.

In terms of $\Psi$, the stress induced by the suspension is $\alpha \Q$, where $\Q(\x,t) = \langle\p\p\rangle$ is the second moment of the distribution function and $\alpha$ is the dipole coefficient, which for $\alpha < 0$ describes extensile dipoles and $\alpha > 0$ describes contractile dipoles \cite{SS2008}. Owing to the small length scales under consideration, the extra stress balances the Stokes equation,
\begin{gather}
-\mu\Delta\u + \grad q = \alpha \grad\cdot\Q + \f,\label{eq:stokes}\\
\grad\cdot\u = 0,\label{eq:divu}
\end{gather}
where $\mu$ is viscosity, $q(\x,t)$ is the pressure, and $\f(\x,t)$ is an external force which, as will be discussed below, arises from fiber elasticity.


The kinetic model, having both spatial and orientational degrees of freedom, is high-dimensional and expensive to simulate. Instead, we reformulate the dynamics in terms of the concentration $c(\x,t) = \langle 1 \rangle$ and second moment tensor $\Q$. This approach eliminates the orientational degrees of freedom of the kinetic theory, but relies on closure assumptions. Here we use a quasiequilibrium closure approximation, commonly known as the Bingham closure, under which the distribution function assumes the form of a maximum entropy distribution,
\begin{equation}
\Psi_B = Z^{-1}(\x,t)e^{\B(\x,t):\p\p},\label{eq:PsiB}
\end{equation}
where the matrix $\B$ and normalization factor $Z$ are determined from the constraints
\begin{align}
    c(\x,t) &= \int_S \Psi_B ~ d\p,\\
    \Q(\x,t) &= \int_S \p\p\Psi_B ~ d\p.
\end{align}
{For a pair of rank-two tensors, the colon operator denotes a double dot product, $\B:\p\p = B_{ij}p_ip_j$.} This closure model exhibits excellent agreement with the kinetic theory for both active and passive suspensions, capturing relevant steady states and their stability and preserving the balance of conformational entropy fluctuations \cite{CL1998,Yu2010,GBJS2017,WSS2022a,Weady2022b}. Integrating Eq. (\ref{eq:dPsi/dt}) against $1$ and $\p\p$ over the unit sphere and assuming a distribution function of the form (\ref{eq:PsiB}), we find
\begin{gather}
\frac{\partial c}{\partial t} + \u\cdot\grad c = D_T\Delta c,\label{eq:dc/dt}\\
\Q^\grad + 2\S_B:\E = D_T\Delta\Q - 6D_R(\Q-c\I/3),\label{eq:dQ/dt_B}
\end{gather}
where $\Q^\grad = \partial\Q/\partial t + \u\cdot\grad\Q - (\grad\u\cdot\Q + \Q\cdot\grad\u^T)$ is the upper convected time derivative, $\E = (\grad\u+\grad\u^T)/2$ is the symmetric rate of strain tensor, and $\S_B = \langle\p\p\p\p\rangle_B$ is the fourth moment of the quasiequilibrium distribution (\ref{eq:PsiB}). {Here the colon operator for a rank-four and rank-two tensor denotes contraction against the last two indices, $(\S_B:\E)_{ij} = S_{B,ijk\ell}E_{k\ell}$.}

On the surface of the sphere $\partial\Omega$, the velocity is no slip $\u\rvert_{\partial\Omega} = 0$ and we use an anchoring condition on the distribution function, $\Psi\rvert_{\partial\Omega} = Z^{-1} e^{\lambda(\hat\n\cdot\p)^2}$. The corresponding boundary condition on $\Q$ is
\begin{equation}
\Q\rvert_{\partial\Omega} = c[\nu\hat\n\hat\n + (1-\nu/2)(\I-\hat\n\hat\n)],\label{eq:Qb}
\end{equation}
where $\hat\n$ is the normal vector and $\nu\in[0,1]$ interpolates between tangential and normal anchoring with $\lambda = \lambda(\nu)$. For simplicity we assume the dipole concentration is initially uniform, $c(\x,0) = N/|\Omega|$, in which case Eq. (\ref{eq:dc/dt}) implies it remains uniform for all time under the boundary condition $c\rvert_{\partial\Omega} = N/|\Omega|$.

\subsection{The polymer model and coupling to the fluid}

We consider a long polymer immersed in the active fluid, which we model as a continuous, flexible fiber. Letting $\X(s,t)$ be a parameterization of the fiber centerline, the elastic energy associated with the fiber is
\begin{equation}
\cE[\X] = \frac{1}{2}\int_0^L E |\X_{ss}|^2 + K(|\X_s|-1)^2 ~ ds,\label{eq:energy1}
\end{equation}
and the fiber moves with the Lagrangian flow map
\begin{equation}
    \frac{\partial\X}{\partial t} = \u(\X,t).\label{eq:dX/dt}
\end{equation}
Here $L$ is the total arclength, $E$ is the bending modulus, and $K$ is the stretching modulus. The fiber ends are taken to be free, in which case we have the boundary conditions $\X_{ss} = \X_{sss} = 0$ at $s = 0,L$.

An additional contribution to the energy arises from alignment interactions with the dipole suspension which may arise from, for example, chemical binding of molecular motors. {For motors that move tangent to the fiber,} such interactions can be captured by the Maier-Saupe potential \cite{MS1958},
\begin{equation}
\cK(\p,\p') = -(\beta/2)(\p\cdot\p')^2,\label{eq:M-S}
\end{equation}
which is commonly used to model interparticle steric interactions \cite{Doi1986,Zhou2007,ESS2013} and is the microscopic analog to the anchoring free energy proposed in Ref. \cite{Kikuchi2009}. This interaction potential is maximized when $\p\perp\p' = 0$, and minimized when $\p\parallel\p'$, where $\p$ and $\p'$ are orientation coordinates such as particle orientation and the fiber tangent vector. {Note that alignment normal to the fiber can be modeled in an analogous way by taking $\beta < 0$.}

For the traditional case of interparticle steric alignment, this potential is integrated against the particle distribution function $\Psi(\p')$. As an analogy, we introduce a distribution function associated with the fiber, $\Psi_f = \delta(\p-\hat\X_s)$, where $\delta$ is the Dirac delta {and $\hat\X_s = \X_s/|\X_s|$ is the unit tangent in the Lagrangian frame}. The net potential a particle with orientation $\p$ experiences is then
\begin{equation}
\begin{aligned}
    \cU_p &= \gamma_p\int_{S'} \cK(\p,\p') \Psi_f ~ d\p' \\ & = 
    -\frac{\gamma_p\beta}{2}\int_{S'} (\p\cdot\p')^2\delta(\p'-\hat\X_s) ~ d\p'\\ & = 
    -\frac{\beta_p}{2}(\p\cdot\hat\X_s)^2.
\end{aligned}
\end{equation}
In the Eulerian frame, this can be written as
\begin{equation}
    \cU_p = -\frac{\beta_p}{2}(\p\cdot\btau)^2,
\end{equation}
where we use $\btau(\x,t)$ to denote the unit tangent vector. In the above $\gamma_p$ is chosen such that $\beta_p = \gamma_p\beta$ has units of inverse time. The potential the fiber experiences due to the suspension is similarly computed by integrating over the particle distribution function,
\begin{equation}
    \begin{aligned}
        \cU_f &= \gamma_f\int_{S'} \cK(\hat\X_s,\p') \Psi ~ d\p'
        \\ & = 
        -\frac{\gamma_f\beta}{2}\int_{S'}(\hat\X_s\cdot\p')^2\Psi ~ d\p'
        \\ & = 
        -\frac{\beta_f}{2} \Q\rvert_{\X}:\hat\X_s\hat\X_s,
    \end{aligned}
\end{equation}
where as above $\gamma_f$ is chosen so that $\gamma_f\beta = \beta_f$ has units of energy per unit length. The torque acting on the particles takes the form $-\grad_p\cU_p = \beta_p(\I - \p\p)\cdot\btau\btau\cdot\p$, and the contribution to the energy functional of the fiber is
\begin{equation}
    \cE_f = -\frac{\beta_f}{2}\int_0^L \Q\rvert_{\X}:\hat\X_s\hat\X_s ~ ds.
\end{equation}
The orientational flux function then becomes
\begin{equation}
\dot\p = (\I-\p\p)\cdot(\grad\u + \beta_p\btau\btau)\cdot\p,\label{eq:pdot2}
\end{equation}
which gives
\begin{equation}
\begin{aligned}
\Q^\grad + 2\S_B:\E &= \beta_p(\btau\btau\cdot\Q + \Q\cdot\btau\btau - 2\S_B:\btau\btau) + D_T\Delta\Q - 6D_R(\Q-c\I/3).\label{eq:dQ/dt}
\end{aligned}
\end{equation}
Similarly, the energy is modified as
\begin{equation}
\cE[\X] = \frac{1}{2}\int_0^L E |\X_{ss}|^2 + K(|\X_s|-1)^2 ~ ds - \beta_f \Q:\hat\X_s\hat\X_s ~ ds.\label{eq:energy2}
\end{equation}
Finally, the force along the fiber centerline is given by the principle of least action, 
\begin{equation}
    \F  = -\frac{\delta\cE}{\delta\X} = -E\X_{ssss} + K\partial_s[(|\X_s|-1)\hat\X_s] - \beta_f\partial_s[(\I - \hat\X_s\hat\X_s)\cdot\Q\rvert_\X\cdot\hat\X_s],
\end{equation}
which in turn imparts a force on the fluid,
\begin{equation}
    \f(\x,t) = \int_0^L \F(s)\delta(\x-\X(s)) ~ ds.
\end{equation}
Note that the alignment potential in the suspension also induces a hydrodynamic stress, however this stress can be omitted at low dipole concentrations \cite{ESS2013}. 

Closer analysis of these equations reveals the following phenomenology. From an energy perspective, when $\beta$ is large, minimizing Eq. (\ref{eq:energy2}) requires $\Q:\hat\X_s\hat\X_s$ to be large as well, which occurs when the fiber is aligned with the principal axis of $\Q$. In this limit, when the two are sharply aligned, at the particle level we have $\btau \propto \p$ from which we find the alignment torque $(\I - \p\p)\cdot\btau\btau\cdot\p$ vanishes, thereby fixing the particle orientation along the fiber. From this perspective, the model is analogous to the active polymer model of Ref. \cite{Saintillan2018} in which the dipoles are forced to be located on and tangent to the fiber. Moreover, it provides an interpretation of the alignment energy as an approximation of a tangential anchoring boundary condition on $\Q$.

\subsection{Non-dimensionalization}

The distribution function is normalized by the number density $n = N/|\Omega|$ so that $\int_\Omega\int_S \Psi ~d\p d\x = |\Omega|$, and the equations of motion are non-dimensionalized by the active timescale $t_c = n\mu/|\alpha|$ and sphere radius $R$. The remaining dimensionless parameters are the dipole coefficient $\alpha' = \sign(\alpha)$, translational and rotational diffusivities $D_T' = (t_c/\ell_c^2)D_T$ and $D_R' = t_cD_R$, arclength $L' = L/R$, elastic moduli $E' = (t_c/\mu\ell_c^2)E$ and $K' = (t_c/\mu)K$, and alignment coefficients $\beta_p' = t_c\beta_p$ and $\beta_f' = (nt_c/\mu)\beta_f$. Under this choice of parameters, the Stokes equation becomes
\begin{gather}
    -\Delta'\u' + \grad' q' = \sign(\alpha)\grad'\cdot\Q' + \f',\label{eq:Stokes_nd}\\
    \grad'\cdot\u' = 0,\label{eq:divu_nd}
\end{gather}
and the $\Q$-tensor equation 
\begin{equation}
{\Q'}^{\grad'} + 2\S_B':\E' = \beta_p'(\btau'\btau'\cdot\Q' + \Q'\cdot\btau'\btau' - 2\S_B':\btau'\btau') + D_T'\Delta'\Q' - 6D_R'(\Q'-\I/3).\label{eq:dQ/dt_nd}
\end{equation}
Similarly, the fiber force density is 
\begin{equation}
    \F'  = -E'\X_{ssss}' + K'\partial_s'[(|\X_s'|-1)\hat\X_s'] - \beta_f'\partial_s'[(\I - \hat\X_s'\hat\X_s')\cdot\Q'\rvert_\X'\cdot\hat\X_s'],\label{eq:F_nd}
\end{equation}
and the Lagrangian flow map remains the same,
\begin{equation}
    \frac{\partial\X'}{\partial t'} = \u'(\X',t').\label{eq:dX/dt_nd}
\end{equation}
For the rest of this work we use dimensionless parameters and omit primes, fixing $\beta_p' = \beta_f' = \beta'$, which ensures appropriate scaling with the underlying interaction potential (\ref{eq:M-S}).

\section{Numerical implementation}\label{sec:implementation}

The complete model consists of the continuum equations (\ref{eq:Stokes_nd})-(\ref{eq:dQ/dt_nd}) along with the fiber force density and evolution equations (\ref{eq:F_nd})-(\ref{eq:dX/dt_nd}). {The continuum equations are discretized with second-order finite differences, using an extension method to enforce boundary conditions in the Lagrangian frame \cite{Kallemov2016,Stein2016}, where the extension domain is chosen as a triply-periodic box.} Time-stepping is done using a first-order implicit-explicit scheme. Fluid-structure interactions are computed via the Immersed Boundary Method, which relates the Lagrangian and Eulerian frames through spreading and interpolation operators 
 \begin{align}
 \cS[\Phi](\x) &= \int_0^L \Phi(s) \delta_h(\x - \X(s)) ~ ds,\\
 \cS^*[\phi](s) &= \int_\Omega \phi(\x) \delta_h(\x - \X(s)) ~ d\x,
 \end{align}
where $\delta_h$ is a regularized delta function \cite{Peskin2002}. The choice of $\delta_h$ is important for approximating grid invariance and maintaining regularity. Here we use the six-point delta function proposed in Ref. \cite{Bao2016}, which is supported over six grid points, has three continuous derivatives, and shows strong grid invariance as compared to the traditional 4-point delta function \cite{Peskin2002}. With these spreading and interpolation operators, the force induced by the fiber on the fluid is compactly expressed as $\f = \cS[\F]$ and the Lagrangian flow map is $\partial\X/\partial t = \cS^*[\u]$. Because the velocity field $\u$ is continuous, this formulation prevents self-intersection of the fiber, avoiding the need for {\em ad hoc} repulsive potentials. {The Eulerian tangent field can similarly be expressed as $\btau = \pi r_h^2 \cS[\hat{\X_s}]$, where $r_h$ is the hydrodynamic radius induced by the regularized delta function.}

Rather than discretize the force density $\F$ directly, we first discretize the integral in $\cE$ and differentiate with respect to the discretization points. Letting $\X = (\X_0,\X_1,\ldots,\X_M)$ be the set of $M$ Lagrangian points along the fiber, the first derivative is approximated as $(\X_s)_m \approx (\X_{m+1}-\X_m)/\Delta s$ and the second derivative as $(\X_{ss})_m \approx (\X_{m+1}-2\X_m + \X_{m-1})/\Delta s^2$, where $\Delta s$ is the grid spacing. The values at $\X_{-1}$ and $\X_{M+1}$ are evaluated using a ghost cell approach based on the boundary conditions $\X_{ss} = \X_{sss} = 0$ at $s = 0,L$. The force density is then computed as $F_{m,i} = -\partial\cE/\partial X_{m,i}$ where $i = 1,2,3$ is the spatial index.

The constraint-based implementation of boundary conditions requires the interpolated fields to match their prescribed boundary values. This constraint induces a Lagrange multiplier at each time step which must be solved for simultaneously with the updated fields. Letting $n$ denote values at time $t^n$, the temporally discretized equations are
\begin{gather}
    -\Delta\u^{n+1} + \grad q^{n+1} + \cB[\bm\Lambda^{n+1}_\u] = \sign(\alpha)\grad\cdot\Q^n + \cS[\F^n],\label{eq:Stokes_discrete}\\
    \grad\cdot\u^{n+1} = 0,\\
    \cB^*[\u^{n+1}] = 0,\label{eq:Bu_discrete}
\end{gather}
with
\begin{equation}
\frac{\X^{n+1} - \X^n}{\Delta t} = \cS^*[\u^n].
\end{equation}
Similarly,
\begin{gather}
    (\cI - \kappa\Delta)\Q^{n+1} + \cB[\bm\Lambda_\Q^{n+1}] = \H^n,\label{eq:dQ/dt_discrete}\\
    \cB^*[\Q^{n+1}] = \Q^b,\label{eq:BQ_discrete}
\end{gather}
where $\bm\Lambda_\u^{n+1}$ and $\bm\Lambda_\Q^{n+1}$ are the Lagrange multipliers, $\kappa = D_T\Delta t$, and $\H^n$ includes the remaining terms in Eq. (\ref{eq:dQ/dt_nd}) evaluated at time $t^n$. Here $\cB$ and $\cB^*$ are the spreading and interpolation operators over the boundary $\partial\Omega$,
\begin{align}
    \cB[\Phi](\bm\x) &= \int_{\partial\Omega} \Phi(\bm\alpha)\delta_h(\x - \X(\bm\alpha)) ~ d\bm\alpha,\\
    \cB^*[\phi](\bm\alpha) &= \int_V \phi(\x)\delta_h(\x - \X(\bm\alpha)) ~ d\x,\\
\end{align}
where $\bm\alpha$ is a Lagrangian coordinate on $\partial\Omega$. Note that the interpolation integral is performed over an extended computational domain $V\supseteq\Omega$, which is chosen as a periodic box. This choice of the computational domain allows us to efficiently invert the implicit linear operators in Eqs. (\ref{eq:Stokes_discrete})-(\ref{eq:Bu_discrete}) and (\ref{eq:dQ/dt_discrete})-(\ref{eq:BQ_discrete}) using the Fast Fourier Transform.

In general, the above linear systems are of the form
\begin{equation}
    \begin{pmatrix} \cL & \cB \\ \cB^* & 0 \end{pmatrix}\begin{pmatrix} u^{n+1} \\ \Lambda^{n+1} \end{pmatrix} = \begin{pmatrix} f^n \\ b^{n+1} \end{pmatrix},
\end{equation}
where $\cL$ is a linear operator, $u^{n+1}$ is the solution vector, $\Lambda^{n+1}$ is a Lagrange multiplier, $f^n$ is a right hand side vector, and $b^{n+1}$ are the prescribed boundary values. Systems of this form can be efficiently solved using a Schur complement approach, which results in a smaller system for the Lagrange multiplier:
\begin{equation}
(\cB^*\cL^{-1}\cB)\Lambda^{n+1} = (\cB^*\cL^{-1})f^n - b^{n+1}.
\end{equation}
Once $\Lambda^{n+1}$ is computed, the solution $u$ is determined by solving $\cL u^{n+1} = f^n - \cB\Lambda^{n+1}$. In the hydrodynamic context, the operator $\cB^*\cL^{-1}\cB$ is called the mobility matrix, which, when appropriately discretized, is symmetric positive definite. This operator only needs to be formed once and is comparatively small, consisting of the number of discretization points on the bounding surface $\partial\Omega$. We form this operator for each of the Stokes system (\ref{eq:Stokes_discrete})-(\ref{eq:Bu_discrete}) and $\Q$-tensor system (\ref{eq:dQ/dt_discrete})-(\ref{eq:BQ_discrete}) prior to simulation, and solve the linear systems using a Cholesky decomposition. Note that just as the immersed boundary formulation avoids overlap of the fiber with itself, imposing boundary conditions in the Lagrangian frame also prevents intersection of the fiber with the boundary.

\section{Linear stability analysis}\label{sec:stability}

\begin{figure*}[t!]
\centering
\includegraphics[scale=0.5]{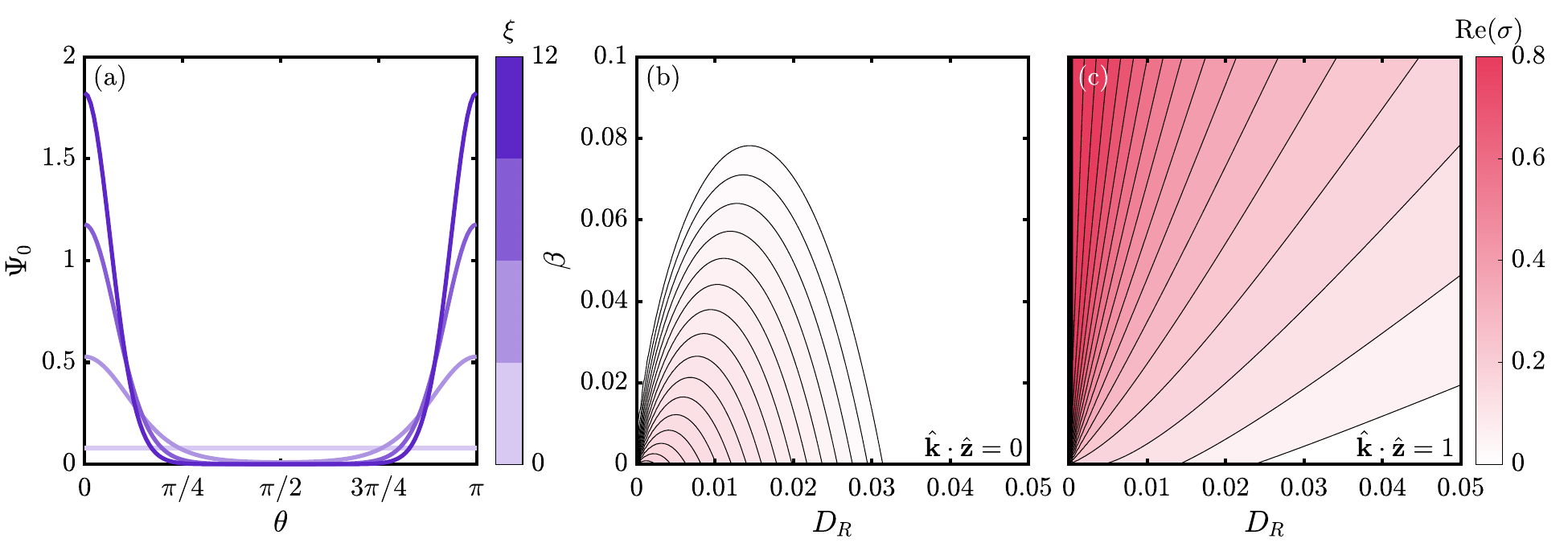}\vspace{-0.125in}
\caption{Linear theory of the coaligned configuration. (a) The bimodal equilibrium distribution, shown for several values of $\xi = 2\beta/D_R$. (b) Growth rate $\text{Re}(\sigma)$ for perturbations with $\hat\k\cdot\hat\z = 0$. For fixed $D_R$, increasing $\beta$ reduces the growth rate, causing such perturbations to decay when $\beta$ is sufficiently large. (c) Growth rate $ \text{Re}(\sigma)$ for perturbations with $\hat\k\cdot\hat\z = 1$. Increasing $\beta$ always has a destabilizing effect, triggering an instability even beyond the threshold of the isotropic state $D_R = 1/32$ for the fluid alone.}\label{fig:stability}
\end{figure*}

To first gain analytical insight into the influence of the fiber on the suspension and the role of alignment interactions, we perform a linear stability analysis of a close-packed fiber in an unconfined domain. In the Eulerian frame, the fiber tangent field $\btau(\x,t)$ satisfies Jeffery's equation \cite{Jeffery1922},
\begin{equation}
\frac{\partial\btau}{\partial t} + \u\cdot\grad\btau = (\I - \btau\btau)\cdot\grad\u\cdot\btau,\label{eq:Jeffery}
\end{equation}
which can be shown by differentiating $\X_s/|\X_s|$ in time and using the Lagrangian flow map (\ref{eq:dX/dt}). The elastic force $\F$ can similarly be translated into the Eulerian frame using the identity ${\partial_s = \btau\cdot\grad}$ under the assumption $|\X_s| = 1$, which will hold to first order in the following analysis. Taking $\btau = \hat\z$ and setting $\dot\p = 0$ yields an equilibrium solution for the distribution function
\begin{equation}\Psi_0 = Z^{-1}e^{\beta\cos^2\theta/2D_R},\label{eq:Psi0}
\end{equation}
where $\theta$ is the polar angle in spherical coordinates and $Z = \int_S e^{\beta\cos^2\theta/2D_R} ~ d\p$. This solution corresponds to a configuration in which the fiber is space-filling and oriented in the vertical direction, and the dipoles are oriented along it with the alignment magnitude set by the ratio $\xi = \beta/2D_R$. The equilibrium distribution is shown in Fig. \ref{fig:stability}(a) and has a bimodal structure, with peaks whose amplitude increases with $\xi$. Note that the case $\xi = 0$ corresponds to an isotropic suspension $\Psi \equiv 1/4\pi$ while $\xi\rightarrow\infty$ corresponds to a sharply aligned suspension.

We consider perturbations about the equilibrium solution of the form $\u = \eps\u'$, $q = \eps q'$, $\Q = \Q_0 + \eps \Q'$, $\btau = \hat\z + \eps \btau'$, where $\Q_0 = \langle\p\p\rangle_0$ is the second moment of the equilibrium distribution (\ref{eq:Psi0}), and $\btau'\cdot\hat\z = 0$ so that $|\btau| = 1$ to $O(\eps^2)$. Taking $E = 0$ and using the fact that $|\X_s| = |\hat\z + \eps\btau'| = 1 + O(\eps^2)$, the only remaining term in the elastic force density is the alignment term $\F = -\beta(\btau\cdot\grad)[(\I - \btau\btau)\cdot\Q\cdot\btau]$. We find the $O(\eps)$ force balance is
\begin{gather}
    -\Delta\u' + \grad q' = \sign(\alpha)\grad\cdot\Q' -\beta (\hat\z\cdot\grad)[(\I - \hat\z\hat\z)\cdot(\Q'\cdot\hat\z + \Q_0\cdot\btau') - (\hat\z\btau' + \btau'\hat\z)\cdot\Q_0\cdot\hat\z],\label{eq:Stokes'}\\\grad\cdot\u' = 0.\label{eq:divu'}
\end{gather}
Similarly, taking $D_T = 0$, the $O(\eps)$ equation for $\Q$ is
\begin{equation}
\begin{aligned}
&\frac{\partial\Q'}{\partial t} - (\grad\u'\cdot\Q_0 + \Q_0\cdot\grad\u'^T) + 2\S_0:\E' \\&= \beta[(\hat\z\btau' + \btau'\hat\z)\cdot\Q_0 + \Q_0\cdot(\hat\z\btau'+\btau'\hat\z) + \hat\z\hat\z\cdot\Q' + \Q'\cdot\hat\z\hat\z -2\S':\hat\z\hat\z - 2\S_0:(\hat\z\btau' + \btau'\hat\z)] - 6D_R\Q',
\end{aligned}\label{eq:dQ'/dt}
\end{equation}
where $\S'$ is the $O(\eps)$ component of the fourth moment tensor. Finally, the $O(\eps)$ tangent field equation is 
\begin{equation}
\frac{\partial \btau'}{\partial t} = (\I - \hat\z\hat\z)\cdot\grad\u'\cdot\hat\z.\label{eq:dtau'/dt}
\end{equation}
In Eq. (\ref{eq:dQ'/dt}), we cannot evaluate $\S'$ from $\Q'$ directly, so for consistency we need to compute it from the perturbation in the exponent $\B = \xi\hat\z\hat\z + \eps\B'$. Linearizing the equilibrium distribution, we get, to $O(\eps^2)$,
\begin{equation}
\begin{aligned}
    \Psi &= \frac{ e^{\xi (\hat\z\cdot\p)^2 + \eps\B':\p\p}}{\int_S  e^{\xi(\hat\z\cdot\p)^2 + \eps\B':\p\p} ~ d\p}
    \\ & \approx
    \frac{e^{\xi(\hat\z\cdot\p)^2}(1 + \eps\B':\p\p)}{\int_S e^{\xi(\hat\z\cdot\p)^2}(1 + \eps\B':\p\p) ~ d\p}
    \\ & \approx
    \Psi_0[1 + \eps(\B':\p\p - \Q_0:\B')].
\end{aligned}
\end{equation}
This implies $\Q' = \S_0:\B' - \Q_0(\Q_0:\B')$, which is a linear system that can be inverted for $\B'$. We then compute $\S' = \T_0:\B' - \S_0(\Q_0:\B')$ where $\T_0 = \langle\p\p\p\p\p\p\rangle_0$ is the sixth moment of the equilibrium distribution. Equations (\ref{eq:Stokes'})-(\ref{eq:dtau'/dt}) form the complete set of linearized equations. 

Introducing plane-wave perturbations in each variable
$(\u',q',\Q',\btau') = (\tu,\tilde q,\tQ,\tilde\btau) e^{\sigma t + i\k\cdot\x}$,
the gradient operator becomes $\grad \mapsto i\k$ and the time derivative $\partial/\partial t \mapsto \sigma$. This yields an eigenvalue problem for the coefficients $(\tilde\u,\tilde q,\tQ,\tilde\btau)$ and eigenvalue $\sigma$ which we solve numerically. Figure \ref{fig:stability} shows the corresponding growth rates $\text{Re}(\sigma)$ in the $(D_R,\beta)$ plane for an extensile suspension ($\sign(\alpha) = -1$); contractile suspensions ($\sign(\alpha) = +1)$ were found to always be linearly stable. We consider perturbations with each (b) $\hat\k\cdot\hat\z = 0$ and (c) $\hat\k\cdot\hat\z = 1$. For the former, alignment interactions always decrease the growth rate compared to the isotropic state, eventually yielding negative values for sufficiently large $\beta$. On the other hand, perturbations of the type in (c) are destabilized by alignment interactions. In particular, for the same parameter values the fiber-fluid system can be linearly unstable even when the active fluid itself would be stable in the absence of the fiber. In all cases the growth rate is independent of the wave amplitude $k = |\k|$, indicating that the dominant wavelength is the system size when spatial diffusion or bending elasticity are included (i.e. $D_T,E > 0$). This type of instability, which is dominated by perturbations such that $\hat\k\cdot\hat\z = 1$, is analogous to the classical bend instability found in dense active suspensions \cite{Simha2002,ESS2013}. In the dilute regime this instability does not occur for an active suspension alone, nor would it, of course, occur for a passive fiber in a passive fluid. Rather, this instability arises uniquely from multiscale interactions at the fluid, particle, and fiber levels.

\section{Numerical simulations}

In this section we simulate the model using the method described in Section \ref{sec:implementation}, considering only extensile suspensions ($\sign(\alpha) = -1$). Contractile suspensions were found to have trivial dynamics for all parameters tested as suggested by linear stability analysis. In all simulations we fix $E = 10^{-4}$ and $K = 10$. {While the bending modulus of chromatin is likely much smaller than that chosen here, we use this value to ensure the fiber is sufficiently resolved under the chosen discretization.} This {small value of $E$} sets a relaxation timescale that is much longer than the active timescale, {while the moderate value of $K$} ensures the fiber maintains unit arclength to approximately $1\%$. We also choose the diffusion coefficients to be $D_T = 5\times10^{-4}$ and $D_R = 10^{-3}$ throughout. For the boundary condition (\ref{eq:Qb}), we set $\nu = 0.1$ so that the suspension is largely tangent to the confining boundary. Though it is an artifact of the computational model, {the hydrodynamic radius $r_h$ can be used to define an effective volume fraction, $\phi = (\pi r_h^2 L)/(4\pi R^3/3)$}. In all simulations we take $\phi = 0.15$, which is comparable to the volume fraction of chromatin during interphase \cite{Dekker2015}. The remaining free parameter is the alignment interaction coefficient $\beta$ whose effects we will study. The fiber is initialized as a random walk inside the sphere with a step size that is uniform in arclength, and is relaxed prior to the simulation by evolving $\partial\X/\partial t = \F - \delta\cE_{\rm ev}/\delta\X$ to ensure the curve is smooth and non-intersecting, where $\cE_{\rm ev} = \sum_{i\neq j}^M 1/|\X_i-\X_j|^2$ is a repulsive potential and the sum is over all pairs of Lagrangian markers along the fiber. {Note that this repulsive potential is only applied during initialization and not during the simulation.}

\begin{figure*}[t!]
\includegraphics{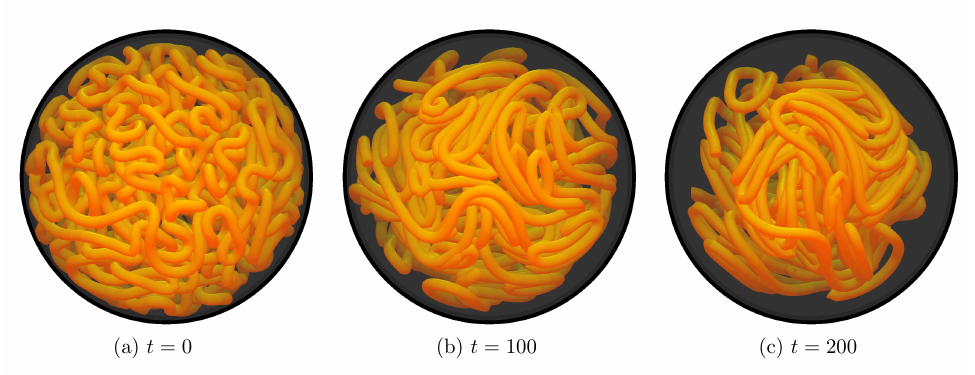}\vspace{-0.125in}
\caption{Snapshots of fiber conformations from a simulation with alignment strength $\beta = 0.1$. The fiber is initially disordered and undergoes persistent fluctuations, gradually condensing towards the center of the sphere. At late times the conformation is characterized by broad regions of high nematic alignment. An accompanying movie can be found as Supplementary Material.}\label{fig:visual}
\end{figure*}

Figure \ref{fig:visual} shows snapshots of fiber conformations from a simulation with $\beta = 0.1$. The fiber undergoes persistent local fluctuations and appears to rotate globally. Over long timescales, the fiber condenses towards the center of the domain and develops elongated segments which are aligned on large scales, similar to what is observed in the active polymer model of Ref. \cite{Saintillan2018}. This latter effect is nontrivial: there are no forces that should produce such alignment {\em a priori}, however interactions with the underlying medium lead to increased order overall. An accompanying movie can be found as Supplementary Material.

\subsection{Large-scale correlations}

\begin{figure*}[t!]
\centering
\includegraphics{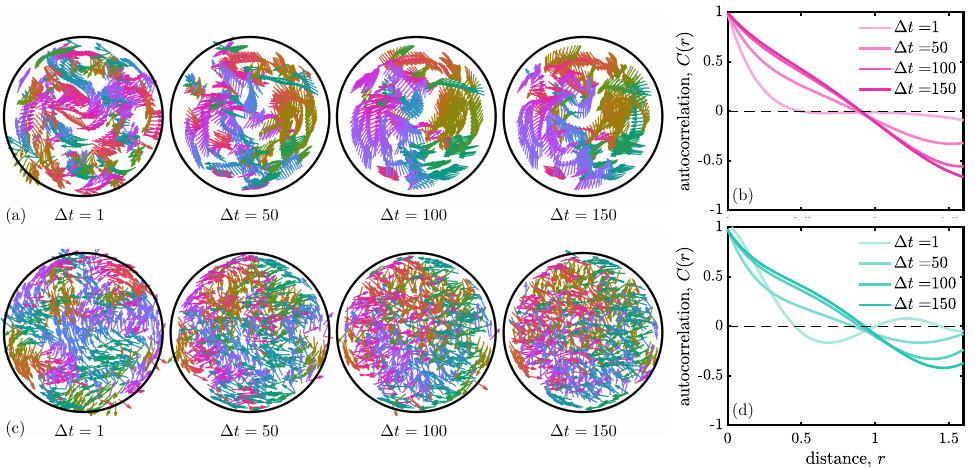}\vspace{-0.125in}
\caption{Large-scale correlations of fiber displacements. (a) The displacement of the fiber shows broad regions of directed motion with increasing time lag $\Delta t$. (b) The spatial autocorrelation function of the displacement map confirms this, showing increased correlations with $\Delta t$ at all scales. At large distances, the autocorrelation function becomes strongly negative, indicating global rotation. (c) Displacement maps of tracer particles without an immersed fiber appear less disordered and remain relatively uniform in space. (d) The autocorrelation function shows correlations increase at intermediate scales with $\Delta t$, but have a slightly lower magnitude.}
\label{fig:displacement}
\end{figure*}

While the fiber dynamics are visually complex, analysis of Lagrangian displacements can help identify coherent motion. Motivated by the analyses of Refs. \cite{Zidovska2013} and \cite{Saintillan2018}, we consider the displacement map
\begin{equation}
\d(s,t;\Delta t) = \X(s,t+\Delta t) - \X(s,t),
\end{equation}
where $\Delta t$ is a time lag. Figure \ref{fig:displacement}(a) shows displacement vectors along a cross section of the sphere at several values of $\Delta t$ for a simulation with $\beta = 0.2$. On short timescales, motion is dominated by fluctuations typical of an active suspension. In fact, because the fiber moves under the Lagrangian flow map, we have $\lim_{\Delta t\rightarrow 0} ~\d(s,t;\Delta t)/\Delta t = \u(\X(s,t),t)$. On long timescales, the displacement vectors are increasingly aligned, reflecting a gradual rotation of the fiber. 

To quantify these observations, we consider the spatial autocorrelation function of the displacement map,
\begin{equation}
C(r) = \frac{\int_0^L\int_0^L \d(s)\cdot\d(s')\delta(|\X(s)-\X(s')| - r) ds ds'}{\int_0^L\int_0^L \delta(|\X(s)-\X(s')| - r) ds ds'}\frac{L}{\int_0^L \d(s)\cdot\d(s) ~ ds}.
\end{equation}
Because the spatial distribution of $\X$ is not necessarily isotropic, the normalization includes the number of points along the fiber at a given distance. Figure \ref{fig:displacement}(b) shows the autocorrelation function of the displacement map at the same times shown in panel (a). We find the correlation amplitude increases at every scale as $\Delta t$ increases, becoming strongly negative as $r$ approaches the size of the domain. This latter feature is characteristic of global rotation.

It is nonetheless unclear whether correlated motion is a consequence of the fiber-suspension coupling or a signature of Lagrangian flow structures of the active suspension. To examine this, we simulate the suspension alone and track passive Lagrangian markers with random initial positions. Figure \ref{fig:displacement}(c) shows displacement maps of the Lagrangian markers using the same parameters as the simulation in Fig. \ref{fig:displacement}(a)-(b). The displacements are visually less structured and appear equally disordered at large time lags. Moreover, the distribution of the Lagrangian markers remains relatively uniform in space whereas the fiber condenses towards the center of the domain. Nonetheless, we find the spatial autocorrelation function of the displacement map, shown in Fig. \ref{fig:displacement}(d), has a similar structure and again increases with the time lag, but exhibits slightly lower correlations overall. In particular, correlations do not grow monotonically at each scale.

\subsection{Conformational dynamics}

The fiber conformation also exhibits a large degree of organization at long times. This structure can be quantified by the fiber nematic tensor
\begin{equation}
\Q_f(\x,t) = \langle \btau\btau - \I/3\rangle_f,
\end{equation}
where $\langle\cdot\rangle_f$ denotes an average over subdomains with radius 1/10 of the sphere diameter. While the total nematic order changes with the size of the subdomain (it will approach one everywhere when the subdomain includes only one point on the fiber), the qualitative behavior is insensitive to this choice.

\begin{figure*}
    \centering
    \includegraphics{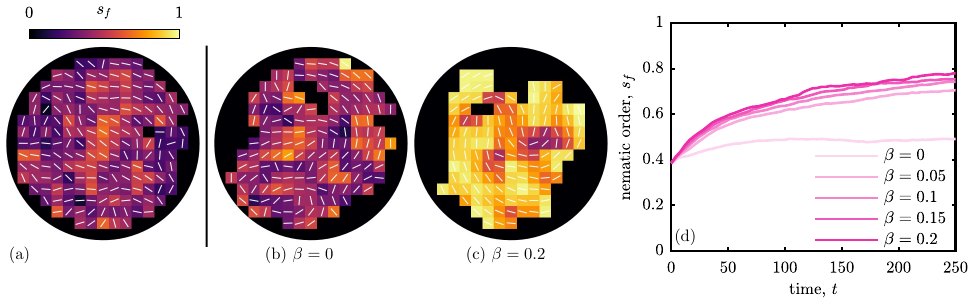}\vspace{-0.125in}
    \caption{Nematic order of the fiber configuration. Panels (a)-(c) show a cross section of the scalar nematic order field $s_f = (3\norm{\Q_f}-1)/2$ with the director $\m_f$ superimposed at a late time.  (a) For the initial configuration, nematic order is relatively low. (b) Order increases only slightly for the case $\beta = 0$. (c) Including alignment interactions, $\beta = 0.2$, increases nematic order, with the fiber exhibiting a large degree of global alignment. (d) The long-time nematic order appears to increase monotonically with $\beta$.}
    \label{fig:fiber-nematic}
\end{figure*}

Figure \ref{fig:fiber-nematic} shows cross sections of the scalar nematic order parameter $s_f(\x,t) = (3\norm{\Q_f}-1)/2$ of the fiber configuration where $\norm{\cdot}$ is the spectral norm (here the largest eigenvalue of $\Q_f$). The director field $\m_f(\x,t)$, which is the principal eigenvector of $\Q_f$, is also superimposed. Panel (a) shows the initial configuration, which has relatively low nematic order as expected from the random initial condition. At later times, for the case $\beta = 0$ shown in panel (b), we find nematic order only increases slightly from the initial configuration and the director remains relatively disordered at large scales. In contrast, the case $\beta = 0.2$, shown in panel (c), shows much higher local nematic order and the director appears to be correlated over large scales. This increased alignment is characterized further by the time evolution of the scalar nematic order parameter, which is shown in Fig. \ref{fig:fiber-nematic}(d) for several values of $\beta$. We find nematic order grows faster with increasing $\beta$ and approaches a larger long-time mean.

\begin{figure*}[t!]
\centering
\includegraphics{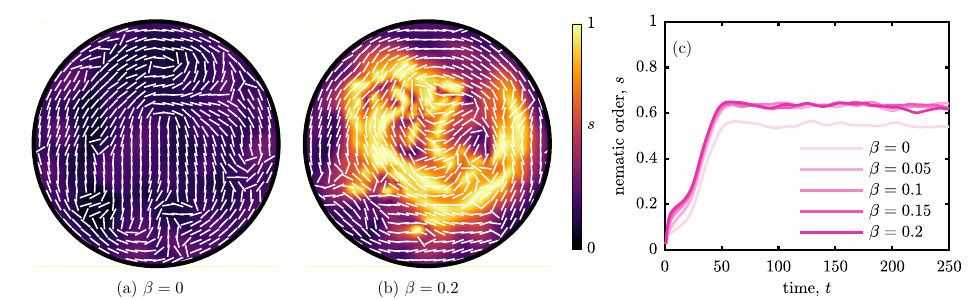}\vspace{-0.125in}
\caption{Nematic order of the dipole suspension. Panels (a) and (b) show cross sections of the scalar nematic order parameter $s = (3\norm{\Q}-1)/2$ for (a) $\beta = 0$ and (b) $\beta = 0.2$ at a late time. The latter exhibits substantially higher nematic order which is organized around the fiber. Panel (c) shows the time evolution of the scalar nematic order parameter for several values of $\beta$, where we find the long-time mean increases with $\beta$.}\label{fig:suspension-nematic}
\end{figure*}

The nematic tensor of the suspension $\Q$ exhibits similar behavior to that of the fiber. Figure \ref{fig:suspension-nematic} shows cross sections of the scalar nematic order parameter $s(\x,t) = 3(\norm{\Q}-1)/2$ for the same simulation in Fig. \ref{fig:fiber-nematic}. We again find nematic order increases significantly with $\beta$, especially in the vicinity of the fiber. Comparing the time evolution of the mean nematic order shows both the growth rate and long-time mean increase with $\beta$, analogous to what is observed for the fiber nematic order. The growth rate of the suspension nematic order, however, is much faster than that of the fiber, reflecting multiple emergent timescales in the system.

\begin{figure}
    \centering
    \includegraphics[scale=0.5]{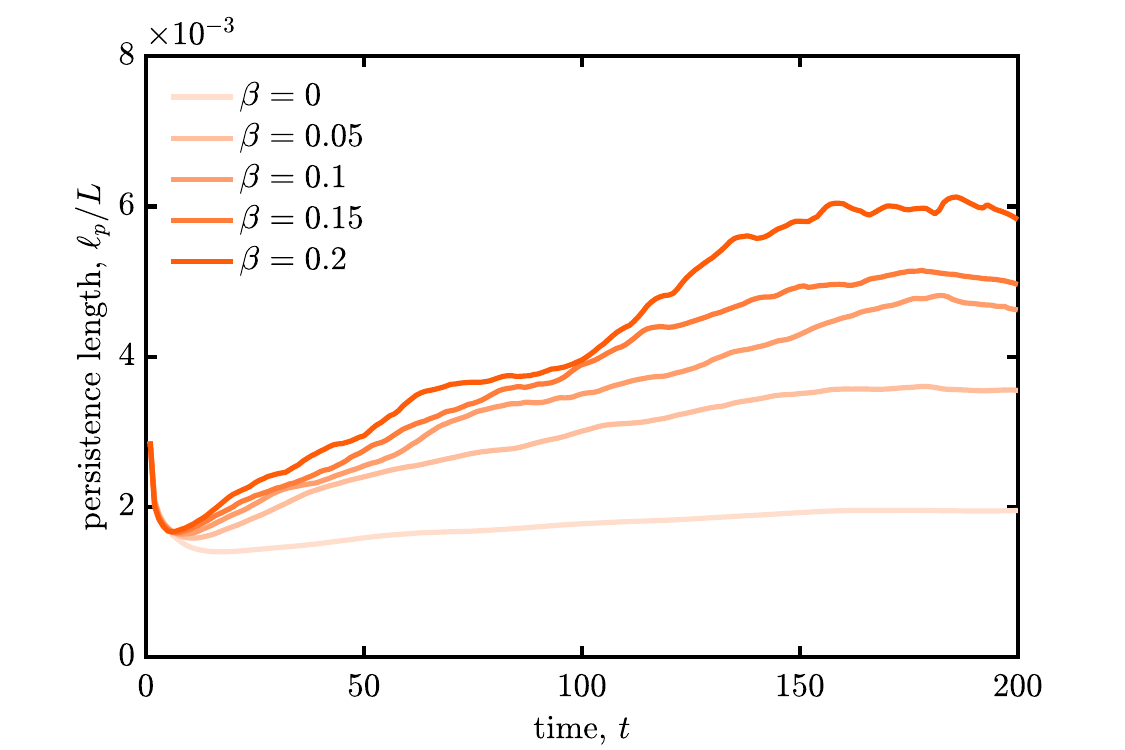}\vspace{-0.125in}
    \caption{Temporal evolution of the effective fiber persistence length, defined in Eq. (\ref{eq:persistence}). Following an initial decay, the persistence length grows over time, with a value that increases with the alignment interaction coefficient $\beta$. For higher alignment strengths, the persistence length tends to show increased fluctuations.}
    \label{fig:persistence}
\end{figure}

Within regions of high nematic order, the fiber elongates and exhibits clear spatial correlations. This elongation is reflective of mechanical rigidity that can be characterized by an effective persistence length $\ell_p$, defined by the relation
\begin{equation}
    e^{-\Delta s/\ell_p} = \langle \hat\X_s(s)\cdot\hat\X_s(s+\Delta s) \rangle_L,\label{eq:persistence}
\end{equation}
where $\langle\cdot\rangle_L$ denotes an average over the fiber centerline \cite{Doi1986}. For a Brownian fiber in equilibrium the persistence length can be computed analytically, and is related to the bending modulus $E$. When immersed in an active medium, this relation does not hold and the persistence length will be modified. (In fact, because there is no noise in our system, the equilibrium persistence length is infinite.)  We compute the effective persistence length from numerical simulations, fitting the parameter $\ell_p$ in Eq. (\ref{eq:persistence}) using five consecutive discretization points along the fiber centerline. Figure \ref{fig:persistence} shows the temporal evolution of the persistence length for simulations with several values of $\beta$. In each case the persistence length sharply decreases at early times, reflecting an instability of the initial configuration, and gradually grows at later times. The growth rate of the persistence length and its long-time mean increase with $\beta$, exhibiting larger fluctuations with larger $\beta$. Similar effects were observed as a function of activity in the active polymer model of Ref. \cite{Saintillan2018}.

\section{Discussion}

This work demonstrates how activity and multiscale fluid-structure interactions can lead to large-scale organization of active processes and passive flexible structures. While highly coarse-grained, the simplicity of the model hints at generic mechanisms that might govern chromatin dynamics. Despite the inverted placement of activity, the phenomenology of our model is strikingly similar to the active polymer model of Ref. \cite{Saintillan2018}. Notably, both models reproduce the activity-driven coherent motions found in live cell nuclei. Indeed an analogy can be drawn between the models for large $\beta$ for which the dipoles are largely tangent to the fiber. This analogy, combined with similarities to the experimental system, further emphasizes the organizing capabilities of active dipoles in Stokes flow, whether bound or unbound, and suggests hydrodynamic interactions play an important role in the genome's dynamic self-organization in the cell nucleus. 

The instability analyzed in Section \ref{sec:stability} is similar to orientational instabilities found in various types of active fluids \cite{Simha2002,ESS2013}. In this case, however, the instability is triggered by interaction with a passive structure rather than higher activity. Moreover, this structure is internal to the fluid and has the opposite effect of rigid confining boundaries, which tend to be stabilizing \cite{Ramaswamy2007,Wioland2013,Chandrakar2020}. Within this analysis, we reformulated the model in a fully continuum setting, yielding a set of equations similar to two-fluid continuum models which have been used to describe chromatin dynamics \cite{Bruinsma2014,Eshghi2021,eshghi2023activity}. This formulation could be studied on its own, including stability analysis in confined geometries, and might inform interpretations of existing continuum theories. 

The model formulation, rooted in microscopic modeling, is versatile and can be extended to include other biologically relevant processes. This may include heterogeneous activity in the fluid to model hetero- and euchromatic regions, multiple fibers representing chromosome territories, binding interactions with the confining boundary such as those induced by lamin proteins \cite{Maji2020}, {or torques induced by twisting of the fiber \cite{Gross2011}}. Moreover, within the immersed boundary framework, the confining boundary can itself be made deformable {to study the effects of nuclear shape fluctuations \cite{Chu2017,Jackson2023}}. While we only considered activity in the fluid, it is straightforward to include activity along the fiber to assess the competing effects of bound and unbound active processes.

\section*{Acknowledgements}

We thank Adam Lamson, Alex Rautu, and David Saintillan for useful discussions. The authors gratefully acknowledge funding from National Science Foundation Grants No. CMMI-1762506 and No. DMS-2153432 (A. Z. and M. J. S.), No. DMR-2004469 (M. J. S.), No. CAREER PHY-1554880 and No. PHY-2210541 (A. Z.).

\bibliography{refs}

\end{document}